\documentstyle[aps,prb,multicol]{revtex}

\begin{document}
\input{epsf}
\draft
\preprint{}
\title{Transport properties of quantum dots with hard walls}
\author{A. Fuhrer, S. L\"uscher, T. Heinzel, and K. Ensslin}
\address{Solid State Physics Laboratory, ETH Z\"{u}rich, 8093
Z\"{u}rich,  Switzerland\\}
\author{W. Wegscheider}
\address{Walter Schottky Institut, TU M\"unchen, 85748 Garching, 
Germany, and\\
Institut f\"ur Angewandte und Experimentelle Physik, Universit\"at 
Regensburg, 93040 Regensburg, Germany\\}
\author{M. Bichler}
\address{Walter Schottky Institut, TU M\"unchen, 85748 Garching, 
Germany}

\date{\today}
\maketitle
\begin{abstract}
Quantum dots are fabricated in a Ga[Al]As-heterostructure by local oxidation 
with an atomic force microscope. This technique, in combination with top gate 
voltages, allows us to generate steep walls at the confining edges and small lateral 
depletion lengths. The confinement is characterized by low-temperature 
magnetotransport measurements, from which the dots' energy spectrum is 
reconstructed. We find that in small dots, the addition spectrum can 
qualitatively be described within a Fock-Darwin model. For a quantitative 
analysis, however, a hard-wall confinement has to be considered. In large dots, 
the energy level spectrum deviates even qualitatively from a Fock-Darwin model. 
The maximum wall steepness achieved is of the order of 0.4 meV/nm.
\end{abstract}
\begin{multicols}{2}
\narrowtext
\section{Introduction}
The electronic transport properties of quantum dots, defined in two-dimensional 
electron gases in semiconductor heterostructures, have received considerable 
theoretical as well as experimental attention. \cite{Kouwenhoven97} 
In the Coulomb blockade regime, where the quantum dot is only weakly coupled 
to reservoirs via tunnel barriers, the conductance through the dot shows 
striking oscillations as a function of a gate voltage which tunes the 
electrochemical potential inside the dot. The separation between these 
so-called Coulomb blockade (CB) resonances contains information not only 
on the single electron charging energy of the dot, but also on its internal 
energy structure. Within the widely used constant-interaction 
approximation,\cite{Beenakker91,Kastner92} the addition spectrum can 
be decomposed into an electrostatic component given by the total capacitance 
of the dot, and a chemical component, which reflects the dot's single 
particle energy spectrum. The Fock-Darwin spectrum\cite{Fock28,Darwin31} 
has been successfully employed in many experiments to describe the energy level 
spectrum of quantum dots.\cite{McEuen91,Staring92,Heinzel95,Tarucha96,Tarucha00} 
These experiments are thus consistent with a harmonic confinement potential 
in two dimensions; to our knowledge, transport signatures of non-parabolic 
confinement in quantum dots have not been reported yet. However, it should be 
emphasized that quantum dots with a hard-wall potential have been fabricated by 
cleaved-edge overgrowth\cite{Schedelbeck97} and by self-assembly\cite{Leonard93}.\\
The option of patterning quantum dots (and other nanostructures) with hard
walls by lithographical means is of interest from both a technological as well as
from a physical point of view. One consequence of steep walls is, for example, 
only a small reduction of the Fermi energy in quantum dots as compared to the 
bulk value. Electrostatic considerations have established a relation between the 
potential steepness and the depletion length.\cite{Larkin95} 
Thus, steeper confinement is a prerequisite for higher pattern densities. 
Also, changes in size and shape are reduced as a gate voltage is tuned. 
This is of particular importance for statistical properties of quantum dots.
\cite{Hackenbroich97} In addition, recent theoretical results on 
conductance fluctuations in quantum dots at small magnetic fields are 
valid only for a hard-wall confinement.\cite{Zozoulenko98} We also
mention the generalized Kohn theorem,\cite{Yip91} which states that in 
parabolic potentials, the far infrared conductivity exhibits only a single 
resonance that corresponds to the characteristic frequency of the bare potential, 
independent of electron-electron interactions.\\
In the present paper, we report the fabrication and characterization 
of quantum dots with super-parabolic confinement potentials. The steepness 
of the walls is increased by combining local oxidation using an atomic force 
microscope (AFM) with an additional top gate electrode. Magnetotransport 
experiments are used to map out the dots' addition spectra, from which the 
energy levels are reconstructed.  The essential idea is to 
detect a super-parabolic confinement via the dots' energy 
spectrum in magnetic fields applied perpendicular to the two-dimensional 
electron gas (Fig. 1): for small Landau level filling factors, 
a parabolic confinement and a hard-wall 
confinement generate quite different energy spectra.\\ 
We have patterned two quantum dots of different size and fabrication 
parameters for the present study: for the smaller dot (labelled ``dot S''),
the spectrum measured is in qualitative agreement with the Fock-Darwin model. 
However, a quantitative analysis reveals inconsistencies, which 
dissolve by assuming a hard-wall confinement. For the larger dot 
(referred to as ``dot L''), the energy spectrum differs \textit{qualitatively} 
from a Fock-Darwin spectrum, and indicates a super-parabolic confinement.\\
It is well known that in the regime of small filling factors, a 
single-particle picture fails to explain the full phenomenology of 
quantum dots, and charge-density model calculations are used instead.
\cite{McEuen92} Therefore, we demonstrate that within a charge-density model, 
our observations can be explained by a steep-wall confinement as well.\\
The paper is organized as follows: in Section II, the theoretical energy 
spectra of  parabolic dots and hard-wall dots are compared. In Section III, 
we describe the sample preparation, the experimental setup, and the electronic 
characterization of two quantum dots. In Section IV, the energy level spectra 
of the dots are reconstructed from magnetotransport measurements and 
the results are discussed. In addition, we estimate the steepness of the walls. 
A summary and conclusion is given in Section V.
\section{Energy spectra of parabolic dots and hard-wall dots}
\subsection{The Fock-Darwin model}
For a circular dot with a parabolic confinement, characterized by the confining strength 
$\omega_{0}$, the energy spectrum is the well-known Fock-Darwin 
spectrum\cite{Fock28,Darwin31} (Fig. 1a):
\begin{equation}
    E_{N,k}=\hbar (N+k)\sqrt{\omega_{0}^{2}+\frac{1}{4}\omega_{C}^{2}}+
    \frac{1}{2}\hbar(N-k-1)\omega_{C}
\end{equation}
Here, we have transformed the radial quantum number m and the angular 
momentum quantum number $\ell$ from the standard representation into 
the Landau level index N (N=1,2,3,\ldots) and the level index 
k of the state within a Landau level (k=0,1,2,\ldots), $k=(m+\frac{|\ell| 
+\ell}{2})$ and $N=(m+\frac{|\ell|-\ell}{2}+1)$.\cite{Hawrylak98} 
Furthermore, $\omega_{C}$ denotes the cyclotron frequency, and each 
level is assumed to be two-fold spin-degenerate. Throughout the paper, 
we restrict ourselves to magnetic fields in which only two spin-degenerate 
Landau levels are occupied, and label the spin-degenerate Landau level 
N as LL(N), N = 1,2. Consequently, the spin-resolved filling factor $\nu$ is always 
in the regime 2 $<\nu<$ 4.\\
A corresponding section of the Fock-Darwin spectrum is 
shown in Fig. 1a. We have chosen typical experimental 
numbers: a dot radius of r = 200 nm, and $\hbar \omega_{0}$ = 1 meV. 
As the magnetic field is tuned, the Fermi level 
varies in zigzag lines, representing the transfer of electrons between 
the two Landau levels. The energies of LL(1)-states drop as B is 
increased, while those of LL(2)-states increase.
A quasi-periodic level crossing between LL(1)-states and 
 LL(2)-states is obtained. Furthermore, the 
density of states is identical in both Landau levels. For 
$\omega_{0} < \frac{1}{2}\omega_{C}$, the separation between 
adjacent states with identical N can be estimated as $\Delta 
E_{N}=E_{N,k+1}-E_{N,k}\approx \hbar\frac{\omega_{0}^{2}}{\omega_{C}}$, and the 
period in B is 
approximated to first order by $\Delta 
B\approx(\frac{\omega_{0}}{\omega_{C}})^{2}B$, as can be 
seen from eq.(1). These approximations are in reasonable agreement 
with $\Delta E_{N}$ and $\Delta B$ in Fig. 1a. 
Note in particular that $\Delta B$ is significantly larger than the 
``bulk value'', which corresponds to the magnetic field needed to change the 
number of magnetic flux quanta through the dot area A by one, $\Delta 
B_{bulk}=\frac{h}{eA}$ = 33 mT. Furthermore, the difference in slope 
between LL(1)-states and LL(2)-states has an upper limit of 
$\frac{dE_{2}}{dB}-\frac{dE_{1}}{dB}=$ 2$\hbar\omega_{C}/B$.
\subsection{The hard-wall potential}
The spectrum of a dot with a hard-wall potential (Fig. 1c) is obtained by 
numerical calculation of the zeroes of the hypergeometric function 
$_{1}F_{1}$, and looks quite different \cite{Geerinckx90}: most strikingly, 
the density of states at the Fermi level in LL(2) 
is higher than in LL(1), provided the Fermi energy is not far above 
$\frac{3}{2}\hbar \omega_{C}$ (Figs. 1c,d). 
Second, $\Delta B$ is well approximated by $\Delta B_{bulk}$. $\Delta 
E_{N}$, however, depends sensitively on N, k and the magnetic field.\\
We note that in the model considerations above, we have neglected spins for clarity. 
Inclusion of the twofold occupation of each orbital state due to 
spin is necessary for a quantitative comparison of the models with the 
experimental data below, and reduces all the above average energy level 
separations and magnetic field periods by a factor of two.\\
\section{Fabrication and characterization of the dots}
\subsection{Sample fabrication by local oxidation}
The samples are patterned out of a shallow Ga[Al]As heterostructure with 
the two-dimensional electron gas (2DEG) $34$ nm below the surface. The 
quantum dots are defined by local oxidation with an AFM.\cite{Held98,Held99} 
The height of the oxide lines is roughly equal to the penetration 
depth of the oxidation into the heterostructure. An oxidation depth of 
approximately 6 nm depletes the 2DEG underneath. We find that the lateral depletion 
length $\ell_{d}$ can be tailored by the oxidation parameters: $\ell_{d}$ 
increases as the height of the oxide line is increased. 
The details of this mechanism are still under investigation.\cite{Scholze2000}
In a simple picture, however, the oxidation can be 
understood as a shallow removal of the semiconductor layers, starting 
at the surface. Removing the oxidized material selectively by a wet etching
step does not change the electronic properties of the AFM defined
nanostructures. We conclude that the patterned surfaces behave pretty
much like a free GaAs (or Al$_{x}$Ga$_{1-x}$As, respectively) surface, 
i.e., the Fermi energy is pinned about mid-gap. Below the oxidized line, 
the sample surface has moved closer to the 2DEG, which
can lead to a depletion of the mobile electrons.
Crucial for the magnitude of the lateral depletion length is the depth of the
oxidation process. In shallow 2DEGs it is well known that most of the
electrons originating from the Si donors charge surface states, while
only a small fraction of them (typically 10 \%)  go 
to the heterointerface and lead to the mobile carriers. The oxidized sample
surface requires additional electrons per area. If the donor layer
remains intact after the oxidation process, these additional electrons
are taken from the underlying 2DEG, which is consequently depleted.
The lateral depletion length is therefore roughly determined by the
distance between the donor layer and the 2DEG.
If, however, the oxidation process penetrates down to the donors (a Si $\delta$-layer 
 located 16 nm below the surface),
they become electronically inactive in this area, and charges from the
neighboring, still intact Si-donors have to compensate the surface
potential in the oxidized regions. Consequently, these electrons are
now missing for the population of the 2DEG in the vicinity of the 
oxide lines, and an increased depletion length results. We estimate that $\ell_{d}$ is
now roughly given by the distance between the 2DEG and the
unpatterned sample surface. This line of thought is sketched in Figs. 
2c, d. Furthermore, the depletion process is supported by the 
 generation of additional surface states at the edges of the oxidized 
 trenches.\\	 
 Dot S is defined by oxide lines with an average height of roughly 20 nm and a 
 width of 100 nm. Its lithographic size is 280 nm$\times$280 nm (Fig. 2a).
 The oxide penetrates about 20 nm into the heterostructure and most likely 
 reaches the donor layer. Dot L has a lithographic area of 400 nm$\times$420 nm. 
 Here, the oxide lines are kept as shallow as possible (height 10 nm).\\
 After the local oxidation, the samples are covered with a homogeneous top gate 
 (tg) electrode. In Figs. 2c and d, schematic cross sections through 
 the dots are shown.
\subsection{Experimental setup and electronic characterization} 
 The measurements were carried out in the mixing chamber of a
 $^{3}$He/$^{4}$He-dilution refrigerator with a base temperature of
 $90\,mK$. The electron density of the ungated two-dimensional electron gas 
 (2DEG) was 5.5 $\cdot$ 10$^{15}$m$^{-2}$, and its mobility was 90 m$^{2}$/Vs. DC bias 
 voltages of 20 $\mu$V were applied across the dot from source to drain, and the current 
 was measured with a resolution of 500 fA. The measurements were
 performed in the weak coupling regime, where the quantum point 
 contacts are adjusted to the tunneling regime by voltages applied to 
 the qpc gates in Figs. 2a,b. The conductance is measured as a 
 function of $V_{I}$, the gate voltage applied to gate I, and Coulomb 
 blockade (CB) oscillations are observed (Figs. 2e,f). Fits of single 
 CB resonances to the lineshape expected for coherent single-level 
 transport,\cite{Beenakker91} i.e., $G\propto \lbrack 4k_{B}T 
 cosh^{2}(\frac{E-E_{r}}{2k_{B}T})\rbrack ^{-1}$,
 reveal an electron temperature of $T_{e}$ = 140 mK ($E_{r}$ denotes 
 the resonance frequency). Magnetic fields up 
 to B = 12 T could be applied perpendicular to the 2DEG.\\
 From measurements of the CB diamonds,\cite{Kouwenhoven97} 
 we determine the single electron 
 charging energies of $E^{S}_{C}=e^2 \slash C_{\Sigma}^{S}$ = 1.22 
 meV for dot S, and $E^{L}_{C}$ = 180 $\mu$eV for dot L.
 By measuring the Coulomb blockade period as a function of the top 
 gate voltage and using the parallel plate capacitor expression, dot S can be modelled 
 as a circular disc with a radius of r $\approx$ 90 nm, which 
 corresponds to $\ell_{d}$ $\approx$ 50 nm at $V_{tg}$ = 
 +100 mV, the working point for this device. The average 
 single-particle energy level spacing can thus be estimated to 
 $\Delta_{S}$ = 140 $\mu$eV. The bulk 
 electron density for this top gate voltage was $n_{2DEG}=5.9\cdot 
 10^{15}m^{-2}$. At B = 8.8 T, 
 the second Landau level is depleted inside the dot, which manifests 
 itself in a sudden transition of the phase diagram structure (not 
 shown)\cite{McEuen92}. Thus, the electron density inside the dot is 
 reduced by $\approx 20\%$, compared to the two-dimensional value.\\
 For dot L, the single electron charging energy is $E^{L}_{C}$ = 180 $\mu$eV,
 and from the capacitance between the top gate and the dot, we deduce
 an electronic dot area of $\approx$ 400 nm $\cdot$ 400 nm. 
 The shallow oxidation, in combination with a large 
 positive top gate voltage of +390 mV (such that the second 
 two-dimensional subband is still empty) generates extremely 
 small depletion lengths of $\ell_{d}\approx$ 15 nm \cite{Held99} and 
 is expected to maximize the steepness of the walls (Fig. 2d) as well. Corrspondingly, the single-particle 
 energy level spacing is $\Delta_{S}\approx$ 22 $\mu$eV. At 
 $V_{tg}$ = +390 mV, we measure $n_{2DEG}$ = 6.3 $\cdot$ 10$^{15}$ m$^{-2}$. At B = 
 11 T, LL(2) becomes depleted inside the dot, which means that its electron density 
 is only slightly smaller than $n_{2DEG}$.\\
 The parameters of the two dots are summarized in Table I.\\
  It is well established that in quantum dots in the regime of
 filling factors 2 $\leq \nu \leq$ 4, 
 the conductance as a function of B and a gate voltage shows 
 periodic patterns over wide ranges of magnetic fields and gate 
 voltages. Furthermore, it has been demonstrated that the energy level spectrum of the 
 dot can be reconstructed from such measurements, which yields 
 information on the confining potential.\cite{McEuen92} We therefore characterize 
 our dots by magnetotransport experiments in this regime of filling 
 factors.\\ 
 Figs. 2e,f compare the corresponding measurements for dot S and 
 L. The small dot (Fig. 2e) shows 
 structures  similar to those reported in earlier 
 experiments:\cite{McEuen91,McEuen92} as the magnetic field 
 is changed, the CB resonances move in zigzag lines. Their average 
 separation in gate voltage corresponds to one Coulomb blockade 
 oscillation period. Here, the lever arm $\eta$ = $\frac{dE}{dV_{I}}$ 
 is $\approx$ 0.11 eV/V, and changes with $V_{I}$. We have 
 determined $\eta$ from measurements of the Coulomb blockade diamonds 
 at B = 0.\cite{Kastner92} In regions where 
 the levels move downwards in energy, their amplitude is high, while in 
 regions where they move upwards in energy, their amplitude is 
 strongly suppressed. Regions of high 
conductance occur when a state belonging to LL(1) aligns with the Fermi level
in source and drain. As the magnetic 
field increases, their energy is reduced. On the other hand, 
states belonging to LL(2) will move upwards in energy as B is 
increased, leading to their depopulation. Since the LL(2) states are 
residing in the inner region of the dot (Fig. 1b), their coupling to the leads 
is small, which results in a strongly suppressed peak amplitude.\\
A similar measurement for dot L, Fig. 2f, reveals a different 
 structure. The CB period at $B=0$ in this sample 
 is $\Delta V_{I}$ = 4.2 mV. For dot L, we found $\eta$ = 
 0.043 eV/V. In contrast to dot S, the 
 separation between successive stripes  of high conductance in 
 $V_{I}$ - direction does \textit{not} correspond to one CB period at 
 B=0, but rather to 2.5 CB periods on average. 
It is not straightforward to see how the conductive LL(1) states are 
connected by LL(2) states, i.e., how the zigzag lines run. We will 
discuss this point in detail in the next Section.
Furthermore, there is an overlap between adjacent regions of high 
conductance along the $V_{I}$ - direction, which we attribute
to thermal activation, since in dot L, the level separation is comparable 
to $k_{B}T$ (see Table I).\\
We emphasize that the structures observed in dots S and L are 
characteristic for the whole region of $2 \leq \nu \leq 4$, and change 
only slightly as B is varied.\\
\section{Reconstruction of the energy spectra}
In this Section, we construct the energy level spectra of the two 
dots from the measurements of Figs. 2e and f. While this is 
straightforward for dot S, it requires a more detailed understanding 
of the addition spectrum measured for dot L, which we gather from 
activated transport experiments. The spectra obtained will be compared 
to the model potentials described in Section II, with the spin splitting included.\\
\subsection{Energy spectrum of dot S}
In Fig. 3a, the occupation numbers of LL(1) and LL(2) and the corresponding 
``phase diagram'' are compared to the measurement. Here, a phase is 
given by ($n_{1},n_{2}$), where $n_{i}$ denotes the number of electrons 
in LL(i). The gaps between the zigzag lines in energy (gate voltage) 
 direction correspond to the charging energy $e^{2}/C_{\Sigma}$, 
 plus the separation between the single-particle energy levels inside 
 the dot. From Fig. 3a, 
 we use the lever arm $\eta$ (see Table I) to subtract $E_{C}^{S}$, and obtain bright 
 lines, Fig. 3b, which correspond to the magnetic field dependence of adjacent LL(1) 
 states. The slightly alternating separations reflect the Zeemann 
 splitting. An average spacing between neighboring LL(i) - states of 
 $\Delta_{1}^{S}$ = 310 $\mu$eV, and $\Delta_{2}^{S} \approx$ 400 $\mu$eV is 
 extracted. Note that since the LL(2) states are not directly 
 visible, their position can only be guessed from the point where the 
 bright lines corresponding to the LL(1) states overlap, 
 and $\Delta_{2}^{S}$ can be no more than a rough estimate.\\
 The 
 Fock-Darwin model in the limit of strong magnetic fields states that 
 $\Delta_{1}^{S} =\Delta_{2}^{S}\approx 
 \frac{\hbar}{2}\frac{\omega_{0}^{2}}{\omega_{C}}$ (the factor of 
 $\frac{1}{2}$ takes the spin splitting into account), from which we 
 obtain $\hbar \omega_{0} \approx$ 2.75 meV, an at first sight reasonable, 
 although large value. Further analysis, however, reveals problems 
 with the interpretation in terms of a Fock-Darwin model:\\
 (i) Since $r=90$ nm, we can calculate $E_{F}$ inside the dot 
 from $E_{F} = \frac{1}{2}m^{*}\omega_{0}^{2}r^{2}$, which would result in 
 $E_{F} = 27$ meV, \textit{larger} than in the 2DEG.\\
 (ii) We observe $\Delta$B = 75 mT. Within the Fock-Darwin model in strong magnetic fields, 
 however, $\Delta B \approx 
 \frac{1}{2}(\frac{\omega_{0}}{\omega_{C}})^{2}B\approx$ 180 mT, is 
 expected.\\
 (iii) The slope of the energy levels, $\frac{dE_{i}}{dB}$ (i denotes 
 the LL index) deviates from the Fock-Darwin values: here, 
  $\frac{dE_{2}}{dB}-\frac{dE_{1}}{dB}\approx 2\hbar \omega_{C}/B$, as 
  observed experimentally in previous work.\cite{McEuen91} However, we estimate 
  $\frac{dE_{2}}{dB}-\frac{dE_{1}}{dB}\approx 5\hbar \omega_{C}/B$, 
  which is much too large, even considering the experimental uncertainty 
  in $\frac{dE_{2}}{dB}$.\\
These inconsistencies can be significantly reduced by assuming a hard-wall 
potential. We measure $\frac{1}{2}\Delta B_{bulk} 
\approx$ 80 mT, in reasonable agreement with the 
 $\Delta B_{bulk}$ expected for r = 90 nm. Also, both the 
energy level separation and $\frac{dE_{2}}{dB}-\frac{dE_{1}}{dB}$ can be much larger than in a 
parabolic dot. \\
Hence, although the behavior of dot S is in qualitative agreement with 
a parabolic confinement, a quantitative analysis suggests that its 
confinement resembles a hard-wall potential.
\subsection{Energy spectrum of dot L}
The period in $V_{I}$ of the conductive regionsin dot L 
does not correspond to the Coulomb blockade oscillation period at B=0, 
as it does in dot S. In order to obtain more information on the
states with suppressed conductance, we measured the Coulomb diamonds, 
i.e. the conductance as a function of both $V_{I}$ and the source-drain 
voltage $V_{sd}$, at B = 8T (Fig. 4).    
At \textit{large} bias voltages, i.e. for $|V_{sd}|>200\mu V$, 
an average CB oscillation period of $\Delta V_{I}=e/C_{I}= 3.5$ mV is 
observed, slightly smaller than $\Delta V_{I}$ observed at B=0. As $|V_{sd}|$ 
is reduced, the conductance gets suppressed for 70 $\%$ of the 
CB resonances, corresponding to a ratio between suppressed peaks 
and visible peaks of 2.5:1. The conductance doublets in $V_{g}$-direction, 
Fig. 2f, remain visible at small $|V_{sd}|$. Furthermore, the 
shape of the diamonds with suppressed conductance tends to follow the 
shape of the nearest diamond in which the conductance outside the 
Coulomb blockade is not suppressed, which indicates that the 
activated current flows predominantly via LL(1) states. 
These observations lead us to conclude that the 
density of LL(2) states at the Fermi level is about 2.5 times the 
density of LL(1) states. LL(2) states get occupied when aligned with 
the Fermi energy, but since their coupling to the leads is 
negligible, transport via these states is not measurable, even at 
large $V_{sd}$. Rather, transport occurs via excitation of LL(1) 
states. \\
With this information, we are now able to reconstruct the energy level 
spectrum of dot L by subtracting the $E_{C}$ from the addition spectrum of Fig. 2f.
 The phase diagram is shown in Fig. 3c. Here, we 
 have determined the phase boundaries due to LL(2) states from the 
 activated transport measurements of Fig. 4, concluding that we must 
 cross $\approx$ 2.5 LL(2)-states on average between adjacent LL(1)-states. 
 The zigzag lines obtained 
 are again separated by one Coulomb period in $V_{I}$, and it can be 
 seen that the different structure originates in the fact that most 
 states at the Fermi level couple extremely poorly to the leads. We 
 point out that alternative choices of the phase boundaries are inconsistent not 
 only with the activated transport measurements, but 
 also with other considerations: if we were to assume that the LL(2) states run 
 like the white, dashed line connecting point 1 and 2 in Fig. 3c, 
 we would cross 3.5 LL(2) states on average between adjacent LL(1) 
 states. In addition, all energy levels would run downwards in energy 
 as B is increased, which would lead to an unphysical population of 
 the dot with electrons by increasing B. If we, on the other hand, 
 were to assume that the LL(2) states run parallel to the line between point 1 and 
 3, we would cross equally many LL(2) states as LL(1) states as $V_{I}$ 
 is tuned, in clear contradiction to the activated transport experiments. 
 Also, the slope of the LL(2) states would be 25 meV/T, which is 
 approximately one order of magnitude larger than their maximum slope of 
 $\frac{3}{2}\hbar\omega_{C}/B$, and we would obtain an 
 unphysically small period for the ``Magneto-Coulomb blockade
 oscillations'',\cite{Vaart97} which means that a change in 
 magnetic field of only 10 mT would be sufficient to change the total 
 number of electrons in the dot by one.\\
  Performing the same procedure as before, we 
 reconstruct the energy spectrum of dot L, Fig. 3d. An 
 energy level separation between LL(1) states of $\Delta_{1}^{L}$ = 160 $\mu$eV, 
 and $\Delta_{2}^{L}$ = 60 $\mu$eV for LL(2) is found. Hence, 
 the density of  LL(2) states at the Fermi level is 2.5 times greater 
 than the density of LL(1) states, which indicates a significant deviation from a 
 Fock-Darwin potential. In addition, note that $\Delta B$ = 13 mT = $\frac{1}{2}\Delta 
B_{bulk}$. This is expected for a hard-wall dot, since for each 
additional flux quantum, two LL(1) states are generated.\\
We thus conclude that the energy spectrum of dot L deviates even 
qualitatively from the Fock-Darwin model, but agrees well with a 
hard-wall confinement. In Table I, we compare the 
parameters of the two dots.\\
\subsection{Hard walls and the charge density model}
So far, we have discussed the energy level spectra within a 
 single-particle picture, and found strong evidence for 
 super-parabolic confinement. It is, however, well-known that certain 
 aspects of the energy spectrum of quantum dots cannot be understood 
 in such a simple picture, but are rather interpreted in terms of  a charge-density 
 model.\cite{Chklovskii92} This is particularly true when strong magnetic 
 fields are present, 
 \cite{McEuen92,Vaart97,Vaart94} and therefore raises the question whether 
 our observations hint towards steep 
 walls also within a charge-density model. 
 To answer this issue, we study the depletion of dot L
 along the vertical arrow in Fig. 3c, i.e., as we proceed from point 
 A to point E. For comparison, we first look at the depletion process 
 within the single-particle picture. The corresponding changes in the energies $E_{i}$ of 
 LL(i) are depicted in Fig. 5a. At our starting point A, we have 
 removed an electron from LL(1) into the reservoirs. At point B, a LL(2)-electron 
 is transferred into the leads. This, however, does not lead 
 to significant conductance, since the coupling of the LL(2)-states to 
 the leads is poor. In point B, the energy of LL(2) drops
 by $\Delta_{2}^{L}$ + $E_{C}$, and that one of LL(1) by just 
 $E_{C}$. 
 The same process takes place at points C and D, with the corresponding 
 energy shifts. At point 
 E, the next LL(1)-electron is removed from the dot. Consequently, 
 3 electrons have been removed from LL(2) and one from LL(1). In 
 order to repeat this process cyclically, $\Delta_{1}^{L}$ has to be 
 larger than $\Delta_{2}^{L}$. This is what we have already established 
 above. 
 At increased source - drain bias voltages, we would start to see 
 activated transport via nearby LL(1)-states at some points between A 
 and E. The magnitude of $V_{I}$ 
 at such activated processes is given by the points where the black, dashed 
 lines intersect with the arrow. The separation between the phase boundary and 
 the arrow is a measure of the activation energy necessary. These 
 varying activation energies give rise to the observed 
 ``meta-diamonds'' in Fig. 4.\\
 Within a charge-density model, the measured phase diagram gets a different 
 interpretation:\cite{McEuen92,Chklovskii92,Evans93}  For filling 
 factors $2\leq \nu \leq 4$, 
 LL(1) forms a compressible ring at the dot edge, Fig. 5b. At the Fermi 
 level, LL(2) is spatially separated 
 from this ring by an incompressible region and forms a compressible disk 
 in the dot center; $n_{i}$ in Fig. 3c now denotes the number of electrons in 
the outer ring (i=1) and in the center disc (i=2) . Here, we 
assume that spin splitting does not generate an additional 
electrostatic structure.
The two compressible regions 
are coupled via an intra-dot capacitance $C_{12}$; charging the 
outer ring will now change the electrostatic energy of the inner dot 
by $E_{12}=e^{2}/C_{12}$. If we perform a gate sweep similar to the one indicated by the 
arrow in Fig. 3b, a removal of an electron from the outer ring would 
change the energy of the ring by $E_{C,1}$, and the one of the 
inner dot by $E_{12}$.\cite{Vaart97} Similarly, removing an electron from the inner 
dot changes its energy by $E_{C,2}$, and that one of the ring 
by $E_{12}$, which gives rise to the energy ladder depicted in Fig. 5c. 
Within this picture, $E_{C,i}$, i=1,2 reflect the single electron charging 
energies of the two compressible regions, and hence the 
size, of the outer ring and the inner dot, respectively. Since 
$E_{C,1}>E_{C,2}$, the area of the ring must be smaller 
than the area of the inner dot. This means that the width of the outer 
compressible stripe, $w_{1}$, is small compared to the dot radius.
In previous work, the relation between the width of the compressible 
regions and the potential profile has been well established:\cite{Chklovskii92} 
$w_{1}\propto (dn(x)/dx)^{-1}$, where n(x) is the local electron density and x the distance 
from the edge of the dot. Since $dn(x)/dx$ is proportional to the 
electric field at x, we conclude that also within the 
charge-density model, the potential walls are steep.
\subsection{Estimation of the wall steepness}
While we observe clear signatures of super-parabolic confinement, 
it is not straightforward to determine the actual steepness of the 
walls. We can give no more than an order of magnitude estimate for the edge steepness 
$\frac{dE}{dx}$. According to Sivan and Imry \cite{Sivan1988}, one 
can approximate $\Delta_{1}$ in a hard-wall dot by $\Delta_{1} 
\approx \hbar \omega_{C} \ell_{B}/L$, where $\ell_{B}$ denotes the 
magnetic length and L the dot diameter. Our measurements of 
$\Delta_{1}$ are in good agreement with this expression, which 
indicates that the level separation is not strongly reduced by a 
finite wall steepness. We can estimate the steepness at the Fermi 
level in dot S from the Fock-Darwin model, despite its shortcomings, 
to $\frac{dE}{dx}$ = $m^{*}\omega_{0}^{2}r$, which gives 
$\frac{dE}{dx}\approx$ 0.6 meV/nm.
Furthermore, the charge-density model can be quantified in some more 
detail for dot L.  Since $E_{12}$ corresponds to the separation 
between adjacent conductive regions in Fig. 3c, we estimate 
$C_{12}\approx 60 aF$. We denote the width of 
the outermost compressible stripe by $w_{1}$, and the width of the 
incompressible stripe as $a_{1}$ (see Fig. 5b). The ratio between the area 
of center region and the outer ring is about 2.5, since we have 
2.5 times more levels in the inner region than in the outer 
ring. We use the formula for a planar 
capacitor\cite{Evans93} to 
estimate the ratio $\frac{w_{1}}{a_{1}}\approx \frac{1}{4}exp(C_{12}/4\epsilon 
\epsilon_{0}r)\approx$ 0.5, leading to $w_{1}\approx$ 25 nm and 
$a_{1}\approx$ 50 nm. The Fermi energy drops completely over the 
distance $a_{1}$. Hence, we can estimate 
$\lbrack \frac{dE}{dx}\rbrack_{L}\approx\frac{E_{F}}{a_{1}}\approx$ 400 $\mu$eV/nm.
This is the \textit{average} wall steepness. 
The edge of dot L is thus about one order of magnitude steeper than in 
conventional dots, as estimated from self-consistent charge-density 
functional calculations.\cite{McEuen92} Since the lateral depletion 
length is smaller in dot L than in dot S, one might qualitatively expect 
that the wall steepness is larger. This, however, is not clearly supported by 
our data.\\ 
\section {Summary}
In conclusion, we have demonstrated that by local oxidation of 
semiconductor heterostructures with an atomic force microscope, 
nanostructures with steep walls can be fabricated. Empirically, the height of the 
oxidized lines determines the lateral depletion length. Using quantum dots as an example, 
we have shown via magnetotransport experiments that the confinement can be 
super-parabolic, i.e., it resembles more a hard-wall potential than a 
parabolic potential. The deviations from a parabolic confinement 
become more pronounced as the dot size is increased. 
The average wall steepness is of the order of 0.4 meV/nm. \\ 
We would like to thank T. Ihn an enlightening discussion.
Financial support by the Swiss National Science Foundation is 
gratefully acknowledged.

\begin{table*}
\begin{caption} \\
Characteristics of the two dots. $\Delta$ denotes the average, 
spin-resolved level 
spacing, as estimated from the dot size, while $\Delta_{i}$ is the 
level spacing in Landau level i, as obtained from the reconstruction 
of the energy spectra. Furthermore, $\eta$ = $\frac{dE}{dV_{I}}$ 
denotes the lever arm.\\
\end{caption}
\begin{tabular}{ccc} 
Dot& Small dot (S)&Large dot (L)\\ \hline \hline
n$_{2DEG}$&5.9$\cdot$10$^{-15}$ m$^{-2}$&6.3$\cdot$10$^{-15}$ m$^{-2}$\\ \hline
lithographic size&280 nm $\times$ 280 nm&400 nm $\times$ 420 nm\\ \hline
E$_{C}$=e$^{2}$/C$_{\Sigma}$&1.22 meV& 0.18 meV\\ \hline
electronic dot size& radius: 90 nm&$\approx$ 400 nm $\times$ 400 nm\\ \hline
$\Delta$&140 $\mu$eV&22 $\mu$eV\\ \hline
$\eta$&0.11eV/V&0.043 eV/V\\ \hline
$\Delta_{1}$&310 $\mu$eV&160 $\mu$eV\\ \hline
$\Delta_{2}$&400 $\mu$eV&60 $\mu$eV\\ \hline
\end{tabular}
\end{table*}

\begin{figure}
     \centerline{\epsfxsize=8.0cm \epsfbox{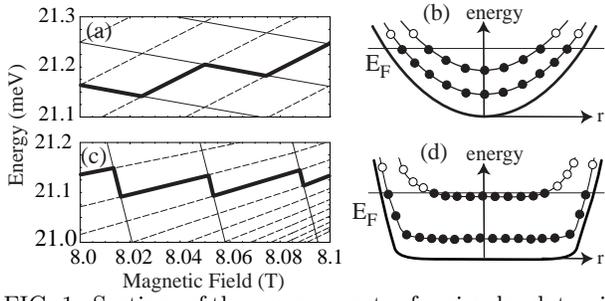}}	
       \caption[Fig. 1] {Sections of the energy spectra for circular 
       dots with a radius of 200 nm and a filling factor $2<\nu<4$, with a parabolic 
       confinement, $\hbar \omega_{0}$ = 1 meV (a) and a hard-wall confinement (c). States 
       belonging to LL(1) (thin full lines) reduce their 
       energy as B is increased, while those states belonging to 
       LL(2) (dashed lines) are running upwards in energy. 
       The bold lines represent the Fermi level when 
       the number of electrons in the dot is constant. In a parabolic 
       dot, the density of states in LL(1) and LL(2) are identical (a,b), 
       while in a hard-wall dot (c,d), the density of states in LL(2) is much 
       larger than in LL(1) at the Fermi level. In (b) and (d), the 
       full circles indicate occupied energy levels, while open 
       circles represent empty states.}
\end{figure}
 \begin{figure}
     \centerline{\epsfxsize=8.0cm \epsfbox{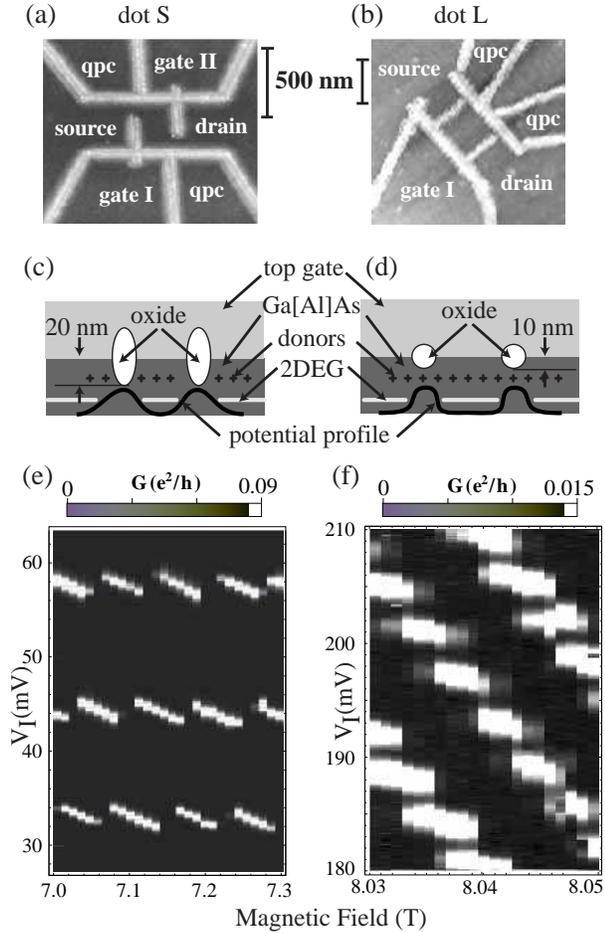}}	
       \caption[Fig. 2] {Surface topography of the two dots (the small dot 
       S, (a), and the large dot L (b)) under study, taken with an 
       AFM before the evaporation of the homogeneous top gate electrode. 
       The bright lines are oxide lines, below which the electron gas is 
       depleted (see text). Both dots are tuned by a voltage applied to
       planar gate I. The gates labelled ``qpc'' are used to adjust the 
       coupling to source and drain. Dot S has a lithographic size of 
       280 nm$\times$280 nm, while the oxide lines are $\approx$ 20 nm in 
       height and 80 nm in width. Dot L is 420 nm$\times$400 nm in lithographic 
       area. Here, the oxide lines are shallower (height 10 nm) and broader 
       (120 nm). Schematic cross sections through dot S and dot L are shown 
       in (c) and (d), respectively, including the top gate. Sketched are also 
       the oxide lines (white ovals) and the resulting potential profile. 
       The 2DEG is 34 nm below the surface. The conductance $G$ of dots S and L 
       at $2 \le \nu \le 4$ as a function of magnetic field and $V_{I}$ are 
       shown in (e) and (f), respectively, in a gray scale plot. The data are 
       taken at a temperature of 90 mK. Both dots show a large conductance 
       (bright areas) only if a LL(1)-state aligns with the Fermi level in the 
       leads (see text).}
     \end{figure}
  \begin{figure}
     \centerline{\epsfxsize=8.0cm \epsfbox{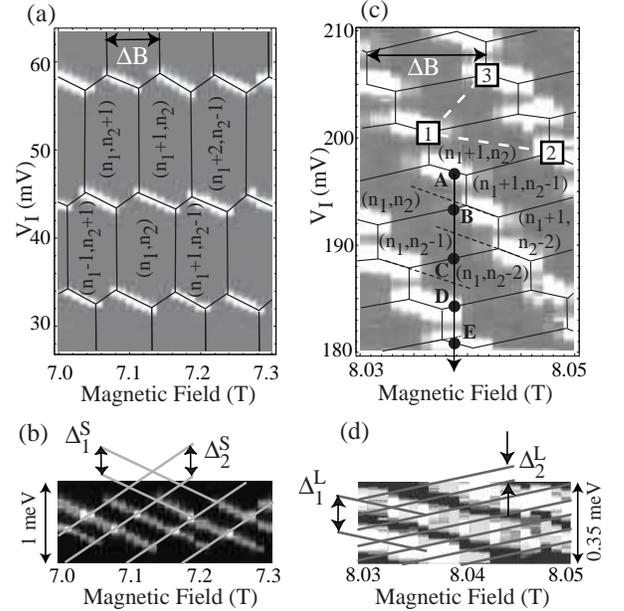}}	
       \caption[Fig. 3] {(a) Sketched phase diagram of the small dot, 
       as an overlay on the data of Fig.1(a).
       Each phase is given by ($n_{1},n_{2}$), where 
       $n_{i}$ denotes the number of electrons in LL(i). If $n_{2}$ 
       changes, the conductance remains zero, due to the poor 
       coupling to the leads. A change in $n_{1}$ results in a high 
       conductance. (b): reconstruction of the energy level 
       spectrum. We find the level 
       spacings $\Delta_{1}^{S}$ = 310 $\mu$eV, $\Delta_{2}^{S}
       \approx$ 400 $\mu$eV. (c) Phase diagram of the large dot, 
       as an overlay on the data of Fig.1(b).
       The phases have the same meaning as in (a). See text on how 
       they were determined. The dark, dashed 
       lines indicate excited LL(1) states. As the capital letters 
       and the bold circles, they refer to Fig. 5 (see text). The 
       bright, dashed lines and the numbers in the white squares denote alternative, 
       but unphysical phase boundaries, and are discussed in the text 
       as well. 
       (d): Reconstruction of the energy level 
       spectrum. The level spacings $\Delta_{1}^{L}$ = 160 $\mu$eV, and 
       $\Delta_{2}^{L} \approx$ 60 $\mu$eV are found.   }
\end{figure}
\begin{figure}
     \centerline{\epsfxsize=8.0cm \epsfbox{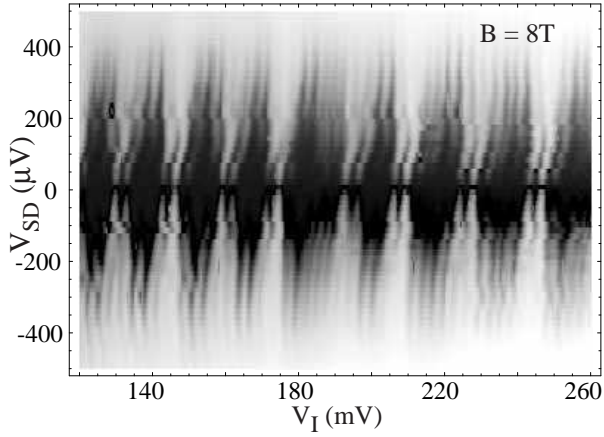}}	
       \caption[Fig. 2] {Coulomb diamonds of the large dot, measured 
       over 40 Coulomb blockade periods, in a gray scale plot. For each 
       completely visible Coulomb 
       diamond, between 2 and 4 diamonds are suppressed, in which the 
       Coulomb blockade oscillations can be 
       observed only under sufficiently large source-drain bias 
       voltages $(|V_{sd}| > 200 \mu V)$. The resulting 
       ``meta-diamonds'' correspond to the energy necessary to carry 
       current via excited LL(1) - states, see text.}
   \end{figure}
  \begin{figure}
      \centerline{\epsfxsize=8.0cm \epsfbox{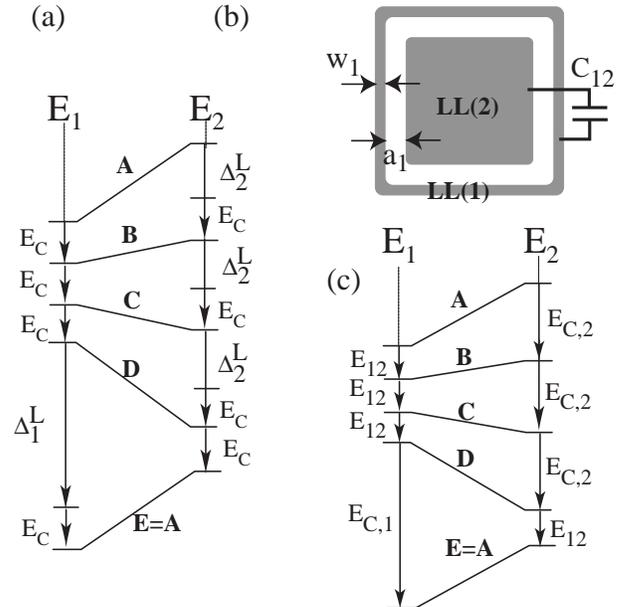}}	
       \caption[Fig. 5] {Energies $E_{1}$ and $E_{2}$ 
       of LL(1) and LL(2) as the large dot 
       is depleted along the arrow in Fig.3(c), in the single- 
       particle picture (a). Within the charge density model, the dot 
       develops an internal electrostatic structure (b). A compressible ring of 
       LL(1) states at the Fermi level 
       is separated by an incompressible region (white) from the 
       compressible region formed by LL(2) states in the center of the 
       dot, and an additional capacitance forms inside the dot. Here, 
       $w_{1}$ and $a_{1}$ denote the width of the outer compressible 
       region and of the incompressible region, respectively. (c): Energies 
       of the core and the ring in the charge density model as the dot is depleted (see text).}
     \end{figure}

\end{multicols}
\end{document}